\documentclass{emulateapj}

\newcommand{\prt}{\partial}
\newcommand{\f}   {\frac}

\begin{document}

\title{Fragmentation of Shocked Flows: 
Gravity, Turbulence and Cooling}

\author{Fabian Heitsch\altaffilmark{1}}
\author{Lee W. Hartmann\altaffilmark{1}}
\author{Andreas Burkert\altaffilmark{2}}
\altaffiltext{1}{Dept. of Astronomy, University of Michigan, 500 Church St., 
                 Ann Arbor, MI 48109-1042, U.S.A}
\altaffiltext{2}{Universit\"ats-Sternwarte M\"unchen, Scheinerstr. 1, 81679 M\"unchen, Germany}
\lefthead{Heitsch et al.}
\righthead{Fragmentation of Shocked Flows}

\begin{abstract}
The observed rapid onset of star formation 
in molecular clouds requires rapid formation of dense fragments which
can collapse individually before being overtaken by global gravitationally-driven
flows.  Many previous investigations have suggested that supersonic turbulence
produces the necessary fragmentation, 
without addressing however the source of this
turbulence. Motivated by our previous (numerical) work on the flow-driven
formation of molecular clouds, we investigate the expected timescales 
of the dynamical and thermal instabilities leading to the rapid fragmentation of gas swept
up by large-scale flows, and compare them with global gravitational
collapse timescales.  We identify parameter regimes in gas density, temperature and 
spatial scale within which a given instability will dominate.
We find that dynamical instabilities disrupt large-scale coherent flows via generation of 
turbulence, while strong thermal fragmentation 
amplifies the resulting low-amplitude density perturbations, thus leading to small-scale, 
high-density fragments as seeds for {\em local} gravity to act upon. 
Global gravity dominates only on the largest scales; large-scale gravitationally-driven
flows promote the formation of groups and clusters of stars formed by
turbulence, thermal fragmentation, and rapid cooling.
\end{abstract}

%
%
\section{Motivation}\label{s:motivation}

There is increasing evidence that star formation (in the solar neighborhood) follows rapidly
on molecular cloud formation (\citealp{2001ApJ...562..852H}; \citealp{2007RMxAA..43..123B};
\citealp{2007ApJ...668.1064E}).
This evidence suggests that the density
enhancements in which stars form are produced {\em during the cloud formation phase} --
and not after the cloud has formed, as implicitly assumed by the initial conditions
of many numerical models. 
Global gravitational modes can sweep up material at the edges of clouds 
\citep{2004ApJ...616..288B}, leading to the formation of filaments 
\citep{2007ApJ...657..870V,2008ApJ...674..316H}. Yet for local gravitational collapse to win, 
the cloud needs to be seeded with local perturbations early on \citep{2004ApJ...616..288B,2007ApJ...654..988H}. 
Not only need these density enhancements to arise early, but they also need to be 
non-linear, i.e. the cloud must acquire strong small-scale density perturbations 
during its formation. Thus, a rapid fragmentation mechanism other -- and faster! -- than gravity is needed.

The scenario of flow-driven cloud formation, in which 
large-scale shocked flows of atomic hydrogen in the warm neutral medium
(WNM) assemble molecular clouds 
(\citealp{1999ApJ...515..286B}; \citealp{2001ApJ...562..852H}), offers an elegant
mechanism for the rapid fragmentation of the proto-cloud and the generation of
non-linear density seeds. The key to the rapid fragmentation is a combination
of strong dynamical and thermal instabilities in shocked flows
(\citealp{1999A&A...351..309H} and \citealp{2000ApJ...532..980K} for one-dimensional 
models; \citealp{2002ApJ...564L..97K} for 2D models including H$_2$ formation; 
\citealp{2005A&A...433....1A}, \citealp{2005ApJ...633L.113H,2006ApJ...648.1052H},
\citealp{2007A&A...465..431H} and \citealp{2007A&A...465..445H}
emphasizing the formation of cold neutral medium clouds in 2D; \citealp{2006ApJ...643..245V},
\citealp{2007ApJ...657..870V} and \citealp{2008ApJ...674..316H} for three-dimensional models, 
the latter two including self-gravity).
Rather than generating turbulence by imposing e.g. a velocity field chosen ad hoc in 
Fourier space, density and velocity structures arise naturally during the formation
of the cloud in this scenario. 

In this study, we attempt to throw some light on the relative importance of
various classes of instabilities at play, via a consideration of their timescales.
We aim to assess whether fragmentation processes other than gravity are acting
rapidly enough during the formation of molecular clouds to allow the rapid onset
of local star formation.
Thermal effects -- strong cooling and thermal instability
(TI, \citealp{1965ApJ...142..531F}) play a dominant role, 
weakening the effective equation of state in a
shocked flow and thus allowing rapid fragmentation. The TI derives its
strength from a combination with turbulence which generates density perturbations 
on small scales. The resulting rapid growth of small-scale perturbations allows the rapid
onset of {\em local} gravitational collapse, before global collapse modes overwhelm 
any pre-existing low-amplitude density variations. Hence, in the context of flow-driven 
cloud formation, thermal fragmentation is the key to the rapid onset of star formation.

The expressions for the timescales and the instability conditions are given in
\S\ref{s:timecond}. We discuss the parameter regimes for the instabilities
and their consequences in \S\ref{s:discussion} and summarize in \S\ref{s:summary}.

%
%
\section{Timescales and Instability Conditions}\label{s:timecond}
The instabilities enabling the rapid fragmentation of shocked flows are
associated with characteristic timescales. These 
will serve as a vehicle to estimate the relative importance
of the instabilities for fragmentation and turbulence generation.

Since we are mainly interested in the competition between global gravity
and local fragmentation processes, 
we restrict ourselves to the discussion of four instabilities which can
be split into two groups of physical processes, namely condensation 
processes (driven by cooling [\S\ref{ss:ti}] and gravity [\S\ref{ss:gi}]) 
and generic fluid instabilities (ram pressure imbalance [\S\ref{ss:NTSI}] 
and shear flows [\S\ref{ss:KHI}]). 
This list of instabilities is by no means exhaustive (\S\ref{ss:oi}).
Our choice is to some extent guided by the results of our numerical simulations
\citep{2008ApJ...674..316H}, and by the notion of cloud formation in large-scale, organized flows 
such as spiral arms of galaxies 
\citep{2003ApJ...599.1157K,2006ApJ...646..213K,2007ApJ...668.1064E,2007MNRAS.376.1747D,2008MNRAS.tmp..285D}, 
or expanding supernova shells (e.g. \citealp{1977ApJ...214..725E}; see also discussion in 
\citealp{2007RMxAA..43..123B}). 

\subsection{Thermal Instability (TI) and Cooling Timescales}\label{ss:ti}
\begin{figure}
  \includegraphics[width=\columnwidth]{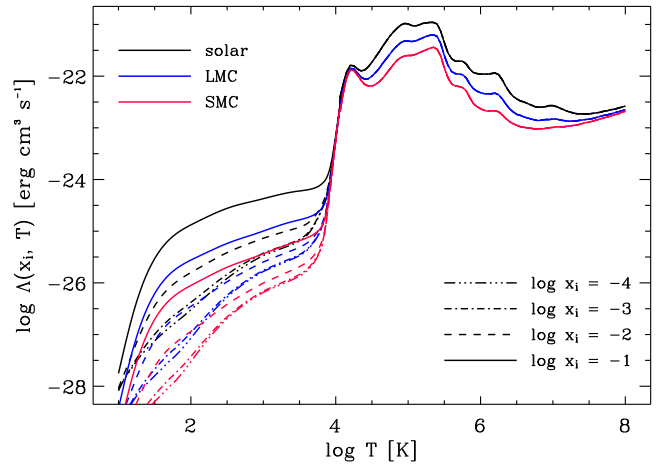}
  \caption{\label{f:coolcurve}Composite cooling curve (see text) 
           for various ionization degrees and for metallicities 
           corresponding to the solar neighborhood, the LMC and the SMC.}
\end{figure}

The TI rests on the balance between heating and cooling processes.
The astrophysically most relevant mode is the isobaric
condensation mode (e.g. \citealp{1995ASPC...80..328B}). 
The condensation mode's linear growth
rate is independent of the wave number as long as the perturbation can
adjust to an isobaric state and under neglection of heat conduction,
however, perturbations on
smaller scales will grow first \citep{2000ApJ...537..270B}. 
Thus, in the linear regime, the condensation mode is limited to scales smaller
than the sound crossing length
(see also \citealp{2003LNP...614..213V} and \citealp{2007A&A...465..431H}), 
\begin{equation}
  \lambda_c = \tau_c\,c_s,\label{e:cond_soundcross}
\end{equation}
where
\begin{equation}
  \tau_c=\frac{k_BT}{n\Lambda}\label{e:ts_cool}
\end{equation}
is the cooling timescale with temperature $T$ and particle density $n$.
Thermal energy losses due to optically thin, collisionally excited atomic lines
are given by the cooling function $\Lambda(T)$ in erg~cm$^3$~s$^{-1}$. The sound
speed is 
\begin{equation}
  c_s\equiv\sqrt\frac{\gamma k_B T}{\mu},\label{e:csound}
\end{equation}
with the mean molecular mass $\mu$.
Since the TI spans a range of 2 orders of magnitude in temperature, 
condition~(\ref{e:cond_soundcross}) can become quite restrictive, but
it varies strongly with temperature and the strength of the cooling 
(see Fig.~\ref{f:coolcurve}).
The scale $\lambda_c$ can drop to a few tenths of a parsec for parameters typical of the WNM.

In the non-linear regime, and
on scales $\lambda>\lambda_c$, the gas can still cool isochorically such that the resulting 
pressure drop will generate waves (e.g. \citealp{1995ASPC...80..328B}), leading to additional
fragmentation.
In that sense, equation~(\ref{e:cond_soundcross}) is not a strict upper limit for the TI, since
it only refers to the (linear) evolution of the condensation mode. 
Nevertheless, we will use it as a proxy for estimating the importance of the TI since it gives
the largest scale out of which a single coherent cold fragment can form and thus is emphasizing
the local nature of the TI. 
On small scales, the condensation mode is limited by the Field (\citeyear{1965ApJ...142..531F})
length, below which heat conduction will become important 
(see e.g. \citealp{2004ApJ...602L..25K}).
The condensation mode will grow if
\begin{equation}
  \left(\f{\prt{\cal L}}{\prt T}\right)_n 
  - \f{n_0}{T_0}\left(\f{\prt{\cal L}}{\prt n}\right)_T<0\label{e:cond_ti},
\end{equation}
where ${\cal L} \equiv n\Lambda-\Gamma$ is the loss-heat function in 
erg~s$^{-1}$ \citep{1965ApJ...142..531F}.

The growth timescale of the condensation mode under the conditions 
(\ref{e:cond_soundcross}) and (\ref{e:cond_ti}) is given by 
\begin{equation}
  \tau_{TI}\equiv\f{\gamma\,k_B\,T}{(\gamma-1)(n\,\prt{\cal L}_n-T\,\prt{\cal L}_T)}
  \label{e:ts_TI}
\end{equation}
with the partial derivatives $\prt{\cal L}_n\equiv \prt{\cal L}/\prt n$ and 
$\prt{\cal L}_T\equiv \prt{\cal L}/\prt T$.
Equation~(\ref{e:ts_TI}) includes the condition for the condensation mode, 
i.e. for $\tau_{TI}<0$, the condensation mode does not grow.

Thermal effects still may be important in the absence of the TI, in which case the
thermal timescale is given by the more general cooling time (eq.~[\ref{e:ts_cool}]). 
A short $\tau_c$ does not necessarily entail TI. 
Rather, a short $\tau_c$ in the presence of a more or less constant heating term
means that the gas cools rapidly to a minimum temperature and then stays at that
temperature effectively isothermal, i.e. the effective adiabatic exponent 
$\gamma_{eff}=1$, as in $P\propto n^{\gamma_{eff}}$. A $\gamma_{eff}<0$ indicates 
thermal instability: with increasing density, the pressure drops. For  
$0<\gamma_{eff}<1$, fragmentation can still be enhanced in the 
presence of an external (gravitational or ram) pressure.

We derive the cooling function from a combination of the rates
quoted by \citet{1972ARA&A..10..375D} and \citet{1995ApJ...443..152W}
for $T<10^4$~K, while for $T>10^4$~K, we use the tabulated curves
of \citet{1993ApJS...88..253S}. 
Figure~\ref{f:coolcurve} summarizes the cooling curves for a range of
ionization degrees $x_i$ and metallicities, corresponding to solar
abundances \citep{1972ARA&A..10..375D} and abundances typical for
the LMC and SMC \citep{1999IAUS..190..266G}. 

\subsection{Gravity, Jeans Instability (GI)\label{ss:gi}}
A (one-dimensional) region becomes gravitationally unstable if 
\begin{equation}
  \rho>\f{(c_s\,k)^2}{4\pi G},\label{e:cond_jeans}
\end{equation}
with the wave number $k$ and mass density $\rho\equiv\mu n$. The corresponding time scale is 
\begin{equation}
  \tau_{GI}=(4\pi G\rho-c_s^2k^2)^{-1/2},\label{e:ts_GI}
\end{equation}
again including the condition for instability. Gravitational collapse occurs for 
$\tau_{GI}\in \mathbb{R}$. 

\subsection{Non-Linear Thin Shell Instability (NTSI)\label{ss:NTSI}}
The NTSI \citep{1994ApJ...428..186V}
arises in a shock-bounded slab, when ripples in a two-dimensional slab
focus incoming shocked material and produce density fluctuations.
The growth rate of the NTSI is mostly controlled by $k\eta$, the product
of the wave number of the slab perturbation $k=2\pi/L$, and the amplitude of the
slab's initial displacement $\eta$ (equivalently, the amplitude of the
collision interface's geometrical perturbation).
The instability is driven by lateral transport of longitudinal momentum,
i.e. if the inflow is parallel to the $x$ direction, and the slab is in the
$y$-$z$-plane, $x$-momentum is transported laterally in $y$ (and $z$),
collecting at the focal points of the perturbed slab. The efficiency of lateral
momentum transport is key to the development of the instability, since
it is the imbalance of ram pressure at the focal points that eventually propels
matter forward, driving the growth of the slab's perturbation.
\citet{1994ApJ...428..186V} derived a growth timescale of
\begin{equation}
  \tau_{NTSI} \approx [c_sk (k\eta)^{1/2}]^{-1}.\label{e:vishniac}
\end{equation}
\citet{1996NewA....1..235B} found that at constant $\eta$ and for small $k$,
equation~(\ref{e:vishniac}) yields only a lower limit, while for large $k$, the
analytical growth rates agree well with the numerical results.
The reason for this seems to lie in the efficiency of deflecting the incoming
flow: for small $k$, a small fraction of the incoming flow's momentum
is converted to lateral motions, while a large part compresses the
slab (depending on the equation of state, this could lead to an increase in energy
losses). 
For the parameter study (\S\ref{s:discussion}), we set $\eta\equiv L/4$.
Keeping $k$ constant and varying $\eta$ affects the NTSI in a similar way as
keeping $\eta$ constant and changing $k$, with a weaker dependence on $\eta$.
Magnetic fields can suppress the NTSI if the ram pressure of the incoming
flow drops below the magnetic pressure \citep{2007ApJ...665..445H}. 
The NTSI is an efficient dynamical focusing mechanism for large-scale gas streams,
resulting in the rapid build-up of massive cores in the focal points 
\citep{2003NewA....8..295H} as possible sites for massive star formation \citep{2008ApJ...674..316H}.

\subsection{Kelvin-Helmholtz-Instability (KHI)}\label{ss:KHI}
The NTSI converts the highly compressible modes of the inflows
into shear flows at the flanks of the perturbed slab. These
shear flows give rise to the KHI, which thus is secondary to
the NTSI, but which also is the main turbulence generation 
mechanism. The turbulence in turn leads to the saturation of the NTSI
\citep{1994ApJ...428..186V}. In the simplest -- incompressible -- scenario 
of the KHI \citep{1961hhs..book.....C}, the shear layer has constant density, and the 
velocity profile is a step function. In this case, the growth timescale is 
given by the velocity difference $\Delta U$ and the wave number as
\begin{equation}
  \tau_{KHI} = (k\Delta U)^{-1},\label{e:ts_KHI}
\end{equation}
i.e. the system is unconditionally unstable. This is definitely the most extreme case: 
magnetic field components parallel to the shear flows can stabilize against the KHI, and 
for compressible flows, the system will be stable for all those wave numbers whose effective
Mach number is larger than a critical value \citep{1968RvMP...40..652G}. For a
detailed discussion of the KHI, see \citet{2008arXiv0802.2497P}.

\subsection{Other Instabilities}\label{ss:oi}
The list of instabilities and their effects considered here is by no means exhaustive. 
For instance, \citet{2006ApJ...652.1331I} discuss a corrugation instability of evaporation fronts
appearing at the interface between the WNM and the CNM on timescales of approximately $\sim 1$~Myr
in the WNM, and down to $0.01$~Myr in the CNM. This -- so they argue -- is on the 
order of the cooling timescales in each of the media, so that the instability
could contribute to the generation of turbulence in the ISM. Hence the corrugation instability
will play an important role for the dynamics of the WNM/CNM. However, as equation~(47) and 
Figure~9 of \citet{2006ApJ...652.1331I} demonstrate, the growth timescale of the corrugation
instability is in the (for our application) interesting range of $\lesssim 1Myr$ only for 
sub-parsec scales, which are substantially smaller than the large-scale flows sweeping up
whole clouds as envisaged in the current study.

In a similar vein, we do not discuss turbulence generation by the TI alone. This possibility 
has been investigated extensively (e.g. \citealp{2000ApJ...537..270B}; \citealp{2002ApJ...569L.127K};
\citealp{2003LNP...614..213V}). \citet{2005ARA&A..43..337C} points out that because of the limited
energy reservoir, thermal pressure variations in a bistable ISM are expected to have negligible dynamical 
effects, requiring additional sources or triggers for turbulence generation (e.g. \citealp{2002ApJ...580L..51K}).

\subsection{Criteria for Instability Dominance}\label{ss:criteria}

The characteristic timescales 
(eqs.~[\ref{e:ts_TI},\ref{e:ts_GI},\ref{e:vishniac},\ref{e:ts_KHI}]) 
can be used to derive criteria for the dominance of a given 
instability in terms of physical quantities. We chose (particle) density and 
temperature, to facilitate a straight-forward comparison to ISM regimes. Attributes 
in square brackets in each of the following subsection titles stand for the line-styles 
used in Figure~\ref{f:ntplane}. Not all combinations are physically relevant. For 
instance, $\tau_c\ll\tau_{GI}$ generally for gravitationally unstable regimes
(\S\ref{ss:gravvstherm}). The ratio of specific heat capacities is set to
 $\gamma\equiv 5/3$, and the mean molecular weight $\equiv 2.36$. Obviously, these
choices are not valid over the full parameter range discussed, but they will considerably
simplify the discussion.

\subsubsection{Gravity threshold [green dashed]}
A region will be gravitationally unstable if 
\begin{eqnarray}
  n &>& \f{\gamma k_B}{4\pi G\mu^2}\,k^2\,T\label{e:gi:rho}\\
    &\approx& 7.4\times 10^1 \frac{T}{[\mbox{K}]}\,\left(\frac{L}{[\mbox{pc}]}\right)^2\,\mbox{cm}^{-3}..
    \nonumber
\end{eqnarray}
This is just equation~(\ref{e:cond_jeans}) expressed in terms of temperature.

\subsubsection{GI dominates NTSI and KHI [red dashed]}
If condition~(\ref{e:gi:rho}) is fulfilled, gravitation will dominate the NTSI if
\begin{eqnarray}
  n &>& \f{\gamma\,k_B}{4\pi G\mu^2}\,k^3\,\eta\,T\label{e:gi_ntsi:rho}\\
    &\approx& 4.7\times10^2 \frac{\eta}{[\mbox{pc}]}\,\left(\frac{L}{[\mbox{pc}]}\right)^{-3}\,
              \frac{T}{[\mbox{K}]}\,\mbox{cm}^{-3},\nonumber
\end{eqnarray}
where $\eta$ is the displacement of the slab. Note the strong dependence on
the wave number $k$, which hints at the power of dynamical fragmentation to
prevent global gravitational collapse.

The condition for the KHI is similar to conditions~(\ref{e:gi:rho}) and (\ref{e:gi_ntsi:rho}). 
The GI dominates over the KHI for
\begin{eqnarray}
  n &>& \f{\gamma\,k_B}{4\pi G\mu^2}\,{\cal M}\,k^2\,T\label{e:gi_khi:rho}\\
    &\approx& 7.4\times 10^1 {\cal M}\,\frac{T}{[\mbox{K}]}\,\left(\frac{L}{[\mbox{pc}]}\right)^{-2}\,\mbox{cm}^{-3},
    \nonumber
\end{eqnarray}
where ${\cal M}$ is the Mach number of the shear flow. This is 
the most beneficial case for the KHI: for supersonic flows,
the KHI is not unconditionally unstable anymore. 

Since conditions~(\ref{e:gi_ntsi:rho}) and (\ref{e:gi_khi:rho}) each depend
linearly on the temperature, they will behave similarly in the $(n,T)$, plane.
Without loss of generality, we can combine them in \S\ref{s:discussion}.

\subsubsection{TI dominates NTSI and KHI [red solid]}
The TI dominates the NTSI for
\begin{eqnarray}
  n &>&\left(\f{\gamma}{\mu}\right)^{1/2}\,k^{3/2}\,\eta^{1/2}\,\f{(k_BT)^{3/2}}{\Lambda(T)}\label{e:ti_ntsi:rho}\\
  &\approx& 5.4\times 10^{-4}\left(\frac{\eta}{[\mbox{pc}]}\right)^{1/2}\left(\frac{L}{[\mbox{pc}]}\right)^{-3/2}
  \left(\frac{T}{[\mbox{K}]}\right)^{3/2}\nonumber\\
  &\times&\left(\frac{10^{26}\Lambda(T)}{[\mbox{erg cm$^3$ s$^{-1}$}]}\right)^{-1}
  \,\mbox{cm}^{-3}.
  \nonumber
\end{eqnarray}
This just mirrors the fact that with increasing density the cooling becomes stronger.
A similar condition can be derived for the KHI by replacing $(k\eta)^{1/2}$ by the 
velocity difference multiplied by the wave number. 

\subsubsection{Sound crossing time limit for the TI [blue solid]}
As discussed in \S\ref{ss:ti}, we use the sound crossing scale (eq.~\ref{e:cond_soundcross})
for the TI's condensation mode as a proxy to estimate the importance of the TI. It leads
to a density threshold above which
the condensation mode of the TI cannot be excited:
\begin{eqnarray}
  n &<& \left(\f{\gamma}{\mu}\right)^{1/2}\,k\,\f{(k_BT)^{3/2}}{\Lambda(T)}\label{e:ti:rho}\\
  &\approx& 2.2\times 10^{-4}\left(\frac{L}{[\mbox{pc}]}\right)^{-1}
  \left(\frac{T}{[\mbox{K}]}\right)^{3/2}\left(\frac{10^{26}\Lambda(T)}{[\mbox{erg cm$^3$ s$^{-1}$}]}\right)^{-1}.
  \nonumber
\end{eqnarray}
The smaller the scales, the less restrictive is the limit on the density for the
condensation mode. The temperature dependence is given by the detailed
form of the cooling curve.

%
%
\section{Discussion}\label{s:discussion}

\begin{figure*}
\begin{center}
  \includegraphics[width=\textwidth]{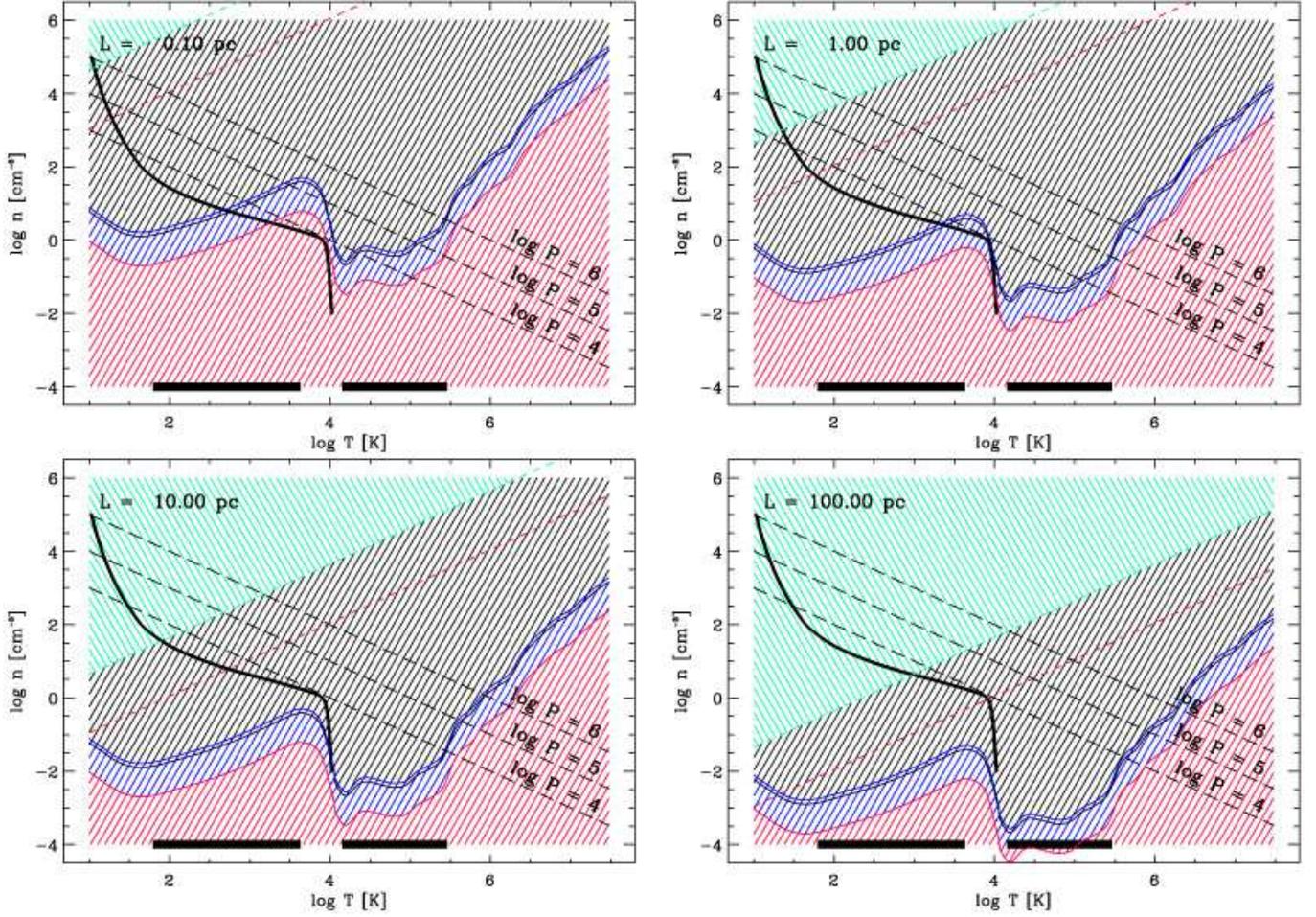}
\end{center}
\caption{\label{f:ntplane}Summary of the instability regimes in 
         $(n,T)$-space, for four
         characteristic spatial scales as indicated in the top left corners.
         The diagonal black long-dashed lines denote lines
         of constant pressure as indicated. The thick black line between $1<\log T<4$
         represents the thermal equilibrium curve due to a balance between heating and  
         cooling. The horizontal thick black lines
         at the bottom of each panel denote the ranges of the two thermally unstable regimes.
         Fragmentation by cooling dominates in all regions with right-slanted hashing (red, blue
         and black), while gravity dominates fragmentation in the left-slanted hashed region (green).
         Red stands for dominance of dynamical instabilities over cooling/TI. Blue/black indicates
         where fragmentation by cooling dominates. The blue ribbon and the black horizontal lines
         together indicate ranges of strict thermal instability.
         See Table~\ref{t:key} for a summary of the line styles and colors.}
\end{figure*}

The conditions summarized in \S\ref{ss:criteria} allow us to 
identify the dominant instabilities in the two-dimensional $(n,T)$ parameter plane.
Most of the conditions (\ref{e:gi:rho})-(\ref{e:ti:rho}) depend on the
spatial scale. Figure~\ref{f:ntplane} summarizes the regimes in four diagrams,
at characteristic (fixed) scales of $0.1,1.0,10$ and $10^2$~pc. 
Table~\ref{t:key} provides a key to the line styles and colors used in 
Figure~\ref{f:ntplane}. We will discuss the dominant instabilities 
(\S\ref{ss:TIdom}--\ref{ss:tivsdi}) and an evolutionary sequence
of a fluid parcel in the ISM (\S\ref{ss:evol}).

\begin{deluxetable*}{llc}
  \tablewidth{0pt}
  \tablecaption{Key to Figure~\ref{f:ntplane}\label{t:key}}
  \tablehead{\colhead{line style}&\colhead{meaning}&\colhead{equation}}
  \startdata
  hashed, slanted right & fragmentation by cooling or TI & \\
  hashed, slanted left  & fragmentation by gravity & \\
  red hashed & dynamics dominate cooling & \\
  blue hashed & cooling/TI dominates dynamics & \\
  black hashed & cooling (but not TI) dominates dynamics & \\
  \hline
  green dash  & Jeans criterion  & (\ref{e:gi:rho})\\
  red dash    & GI dominates NTSI& (\ref{e:gi_ntsi:rho})\\
  red solid   & TI dominates NTSI& (\ref{e:ti_ntsi:rho})\\
  blue solid  & TI crossing time & (\ref{e:ti:rho})
  \enddata
\end{deluxetable*}

\subsection{Thermally Dominated Regime\label{ss:TIdom}}
The most prominent feature in Figure~\ref{f:ntplane} is the large extent of
the thermally dominated parameter regime (right-slanted hashing; blue and black). 
This regime is limited by the growing importance of dynamical instabilities (red) towards
low densities, and by the Jeans condition (eq.~\ref{e:gi:rho}) towards high densities.
Towards low densities (red), the thermal timescales get longer than the dynamical
timescales, so that the gas will tend to behave adiabatically. 
Within the blue hashed region, cooling dominates dynamical instabilities, leading
to the TI within the temperature ranges indicated by the thick black
horizontal bars. Outside the strictly thermally unstable regions, cooling still
dominates and can lead to fragmentation when an external (ram or gravitational) pressure is applied. 
Thus, while the TI dominates the dynamics only within certain temperature ranges,
thermal effects (strong cooling) continue to dominate all through the black
hashed region at higher densities. The upper edge of the blue ribbon is determined
by the sound crossing time condition~(\ref{e:ti:rho}), which is -- strictly -- only
applicable to the TI. There are 
two thermally unstable regimes (according to eq.~[\ref{e:cond_ti}]): 
The lower one between $10^2<T<10^4$~K connects the warm and cold neutral medium (WNM and CNM), 
while the upper (of lesser strength) in the range of $T\approx 10^5$ connects the hot 
and warm ionized medium (HIM and WIM). 

The curved thick black line between $10<T<10^4$~K denotes the thermal equilibrium curve,
where heating terms balance cooling terms. At $T\approx 10^4$~K and $n<1$~cm$^{-3}$, the 
ISM behaves quasi-isothermal. The signature of the thermal instability is a pressure loss
with increasing density  (compare to dashed lines of constant pressure). Moving to
high densities and lower temperatures, the effective equation of state tends back to 
isothermality but stays sub-isothermal, i.e. $\gamma_{eff}<1$. The thermal equilibrium curve
is describing an approximate evolutionary sequence of a fluid element from the WNM to the CNM
(see \S\ref{ss:evol}).

Before we compare the relative strengths of the instabilities, 
Figure~\ref{f:tidetail} offers a more detailed view of the TI in the WNM. As orientation and
for comparison with Figure~\ref{f:ntplane}, the lines of constant pressure and 
the sound crossing time limit (eq.~[\ref{e:ti:rho}]) for the TI have been indicated in
the same style as in Figure~\ref{f:ntplane}.
The color-shaded region denotes the growth timescale (eq.~[\ref{e:ts_TI}]) of the TI 
(see color bar to the right of the diagrams) in all locations where the TI can be 
excited. There are two points to notice here:

\begin{figure*}
\begin{center}
  \includegraphics[width=\textwidth]{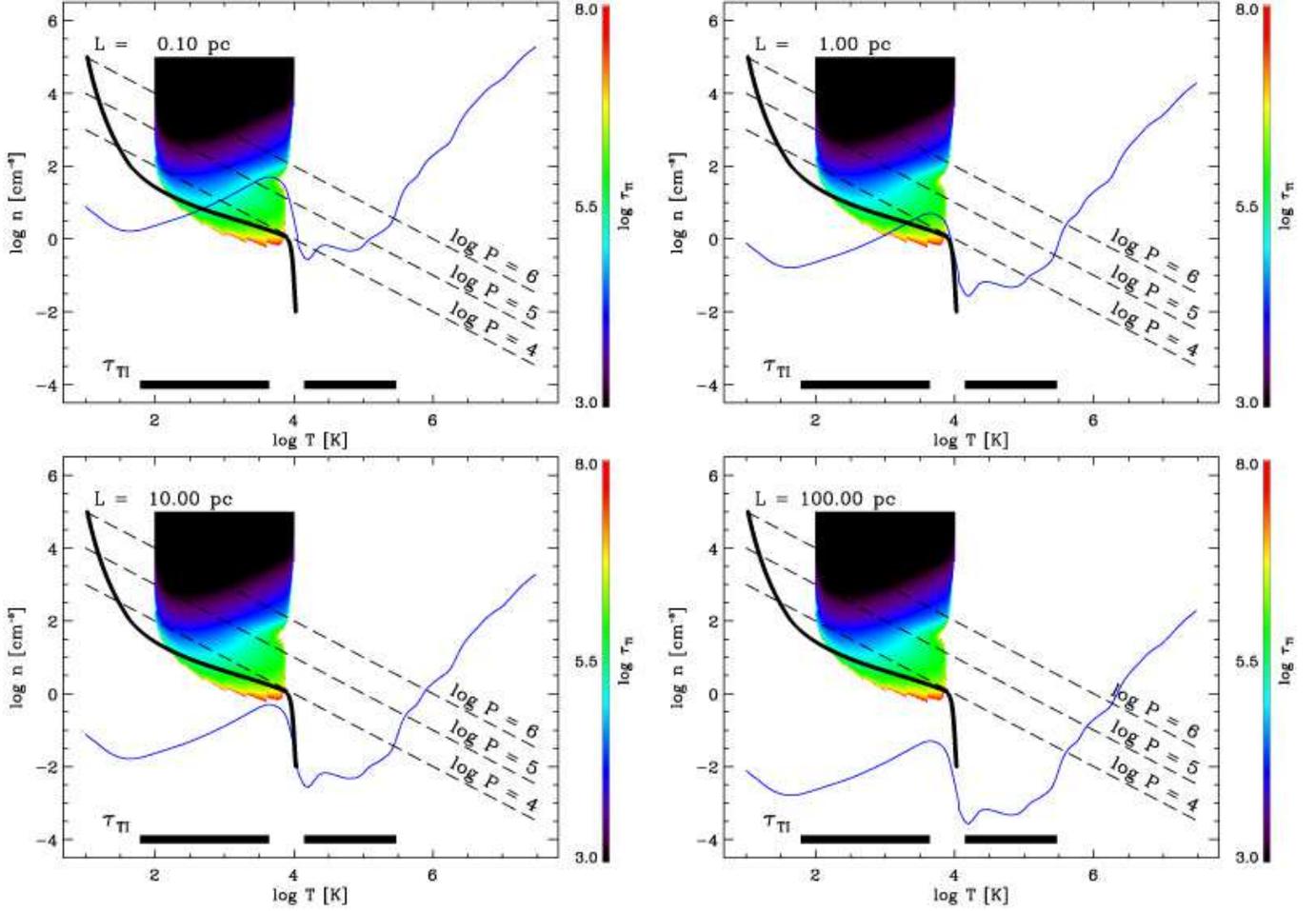}
\end{center}
\caption{\label{f:tidetail}A more detailed look at the TI in 
         $(n,T)$-space, for four characteristic spatial scales as indicated in the top 
         left corners. The diagonal black long-dashed lines denote lines
         of constant pressure as indicated. The thick black line between $1<\log T<4$
         represents the thermal equilibrium curve due to a balance between heating and  
         cooling. The horizontal thick black lines
         at the bottom of each panel denote the ranges of the two thermally unstable regimes.
         The blue line denotes the 
         maximum density given by the sound crossing time limit (eq.~[\ref{e:ti:rho}]).
         The color-shaded region shows the growth timescale of the TI (eq.~[\ref{e:ts_TI}]).
         Light green corresponds to a timescale of $\approx 1$~Myr.}
\end{figure*}

(1) Looking at the upper left panel ($L=0.1$~pc), 
the timescale $\tau_{TI}$ {\em can} become extremely short; however, around the
thermally unstable region, it is on the order of $1$~Myr (light green, traced 
by the thermal equilibrium curve), comparable to dynamical
timescales. This leads to a noticeable amount of thermally unstable gas (e.g.
\citealp{2003ApJ...586.1067H} for observations of neutral hydrogen, and 
\citealp{2001ApJ...557L.121G}, \citealp{2002ApJ...577..768S}, 
\citealp{2005A&A...433....1A}, \citealp{2006ApJ...648.1052H} 
for evidence from numerical models).
For the isothermal branch at $T\approx 10^4$~K, the TI is essentially 
absent. Thus, the lower the density on the (warm) isothermal branch, the higher the
compression due to e.g. shocks needs to be to trigger the TI. Since the post-shock
density in an isothermal gas scales with the square of the Mach number, this condition
is not overly restrictive.

(2) With increasing spatial scale (solid blue line in the remaining three panels of 
Fig.~\ref{f:tidetail}), the upper density limit for the TI drops due to the sound 
crossing time condition~(\ref{e:ti:rho}). Strictly, the TI will only be excited if 
there is a color-shaded region below the blue line, so to speak. This leaves us with 
the somewhat puzzling result that for scales $L>10$~pc, the TI seemingly cannot be excited. 
The solution is two-fold: (a) At large scales (and lower densities/higher temperatures), 
the upper TI at $10^5$~K kicks in, and (b) the notion of fixed scales is misleading. 
The TI will be triggered by compressions and/or turbulent mixing equivalent to a
reduction of spatial scale. Moreover, since it is a condensation mode, 
we cannot regard the evolution of the TI at a fixed scale, but need to follow a 
fluid parcel (see \S\ref{ss:evol}). 

\subsection{Gravitation against Thermal Instability\label{ss:gravvstherm}}
The Jeans criterion (eq.~[\ref{e:gi:rho}]) is fulfilled for all pairs $(n,T)$ above 
the green dashed line in Figure~\ref{f:ntplane} -- e.g., for densities 
$n>10^5$~cm$^{-3}$ and temperatures $T\approx 10$~K at $0.1$~pc. With increasing 
scale, the $(n,T)$-regime dominated by
gravitation extends more and more to lower densities and higher temperatures.
At large scales ($\approx 10$~pc), gravity starts to dominate the parameter space, 
{\em indicating the importance of global gravitational modes}. 

\begin{figure*}
\begin{center}
  \includegraphics[width=\textwidth]{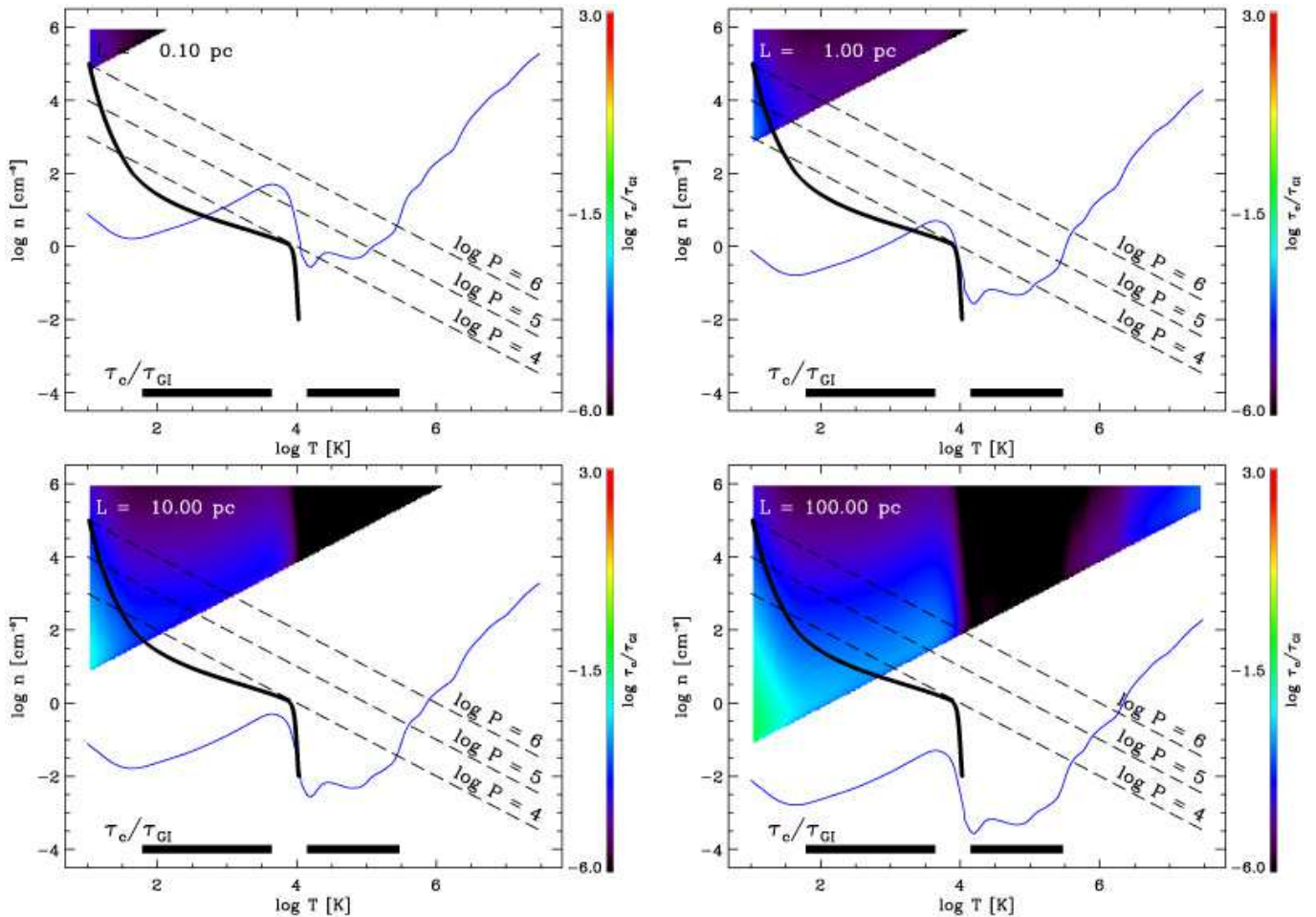}
\end{center}
\caption{\label{f:tipair}Comparison of characteristic times for self-gravity and cooling in 
         $(n,T)$-space, for four
         characteristic spatial scales as indicated in the top left corners.
         The diagonal black long-dashed lines denote lines
         of constant pressure as indicated. The thick black line between $1<\log T<4$
         represents the thermal equilibrium curve due to a balance between heating and  
         cooling. The horizontal thick black lines
         at the bottom of each panel denote the ranges of the two thermally unstable regimes.
         The blue line denotes the 
         maximum density given by the sound crossing time limit (eq.~[\ref{e:ti:rho}]).
         The color-shaded triangles show the ratio $\tau_c/\tau_{GI}$ 
         (eqs.~[\ref{e:ts_cool}], [\ref{e:ts_GI}]) for the pairs of $(n,T)$ 
         subject to the GI. Yellow would correspond to $\tau_c = \tau_{GI}$ -- obviously,
         $\tau_c < \tau_{GI}$ for all gravitationally dominated pairs of $(n,T)$.}
\end{figure*}

Figure~\ref{f:tipair} compares the relative strengths of gravitational over thermal 
condensation\footnote{We are using ``thermal condensation'' for condensation
due to strong cooling, not -- as it occasionally happens in the literature -- for 
gravitational collapse.}. Shown are the same plots as in the previous figures, but 
now overlaid with the ratio $\tau_c/\tau_{GI}$ for the gravitationally dominated 
$(n,T)$-regime (eq.~\ref{e:ts_GI}). Yellow (see color bar of $\tau_c/\tau_{GI}$)
denotes $\tau_c = \tau_{GI}$. The cooling timescales are substantially 
shorter than any gravitational timescales over the whole parameter range accessible 
to the GI. If the TI can be excited (e.g between $80<T<5000$~K), this means that 
{\em thermal condensations will be the dominant fragmentation process}. For parameters 
$(n,T)$ not subjected to the TI, the short cooling timescales mean only that any 
excess energy due to compressions (by gravity!) will be efficiently radiated away. 

\subsection{Gravitational against dynamical effects\label{ss:givsdi}}
The relative importance of gravity and dynamics can be read off 
equations~(\ref{e:gi:rho})--(\ref{e:gi_khi:rho}). The discerning line
for GI against NTSI is shown in Figure~\ref{f:ntplane} as a red dashed line
-- obviously, the NTSI ceases to be important before densities high enough
for gravitational collapse can be reached. Equations~(\ref{e:gi:rho}) and
(\ref{e:gi_khi:rho}) on the other hand show that for Mach numbers ${\cal M}>1$,
the KHI can still dominate in gravitationally unstable gas. This is the regime
of turbulent fragmentation in a Jeans-unstable medium discussed in detail
by \citet{2004RvMP...76..125M}.

\subsection{Thermal against dynamical effects\label{ss:tivsdi}}
Dynamical instabilities dominate below the red (NTSI) and the black (KHI) lines 
(Fig.~\ref{f:ntplane}), generally however in the red hashed region. 
They impose a lower density threshold on the importance of 
cooling (and of the TI within the applicable temperature range), i.e. the gas
will tend to behave more and more adiabatically for those regions in $(n,T)$-space.

\begin{figure*}
\begin{center}
  \includegraphics[width=\textwidth]{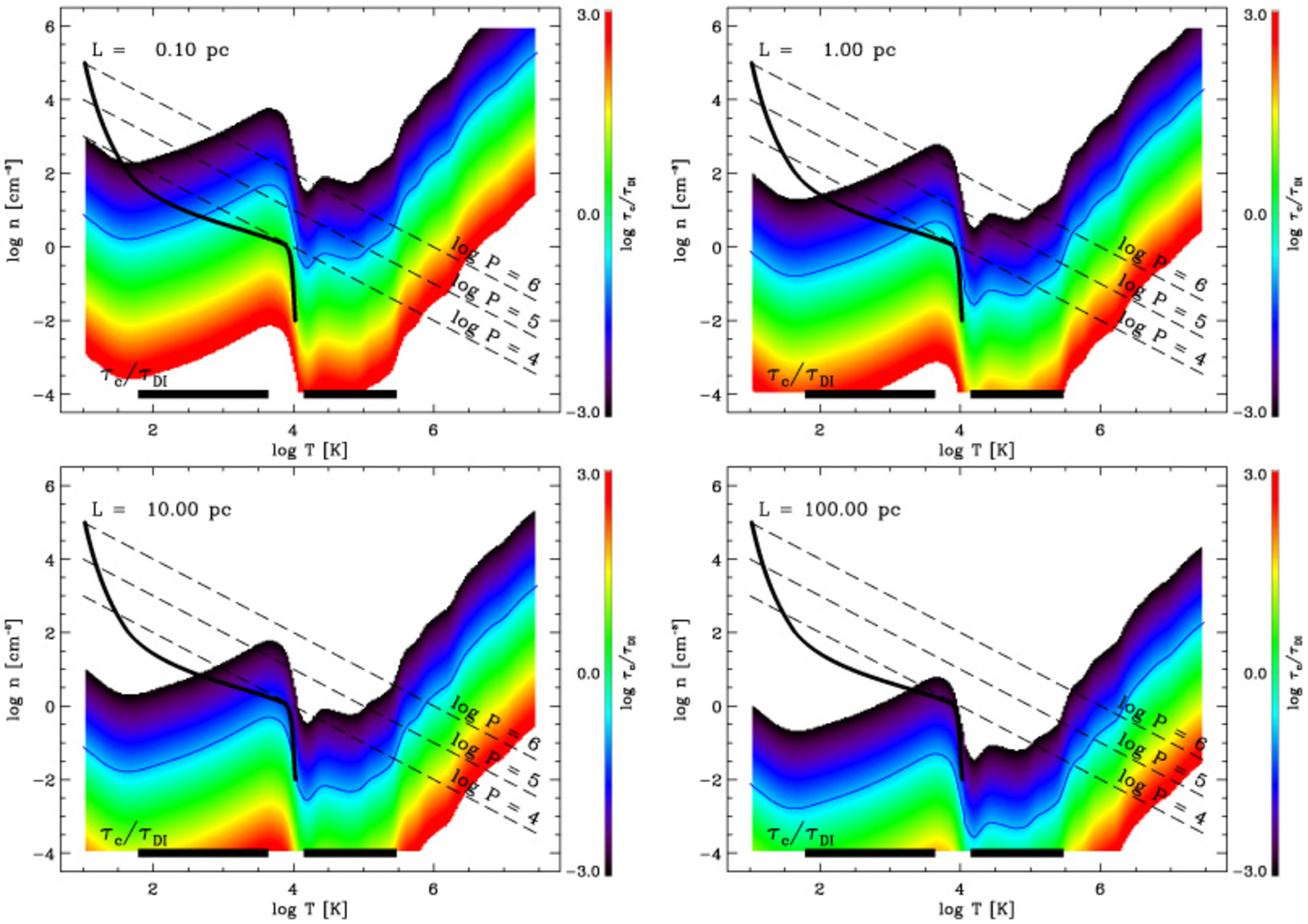}
\end{center}
\caption{\label{f:tidi}Comparison of characteristic times for dynamical instabilities 
         and cooling in $(n,T)$-space, for four
         characteristic spatial scales as indicated in the top left corners.
         The diagonal black long-dashed lines denote lines
         of constant pressure as indicated. The thick black line between $1<\log T<4$
         represents the thermal equilibrium curve due to a balance between heating and  
         cooling. The horizontal thick black lines
         at the bottom of each panel denote the ranges of the two thermally unstable regimes.
         The blue line denotes the 
         maximum density given by the sound crossing time limit (eq.~[\ref{e:ti:rho}]).
         The color-shaded regions show the ratio $\tau_c/\tau_{DI}$ (eqs.~[\ref{e:ts_cool}], 
         [\ref{e:ti_ntsi:rho}]). Note that the ratio decreases monotonically with increasing
         density.}
\end{figure*}

Figure~\ref{f:tidi} shows the ratio of the cooling time scale (eq.~[\ref{e:ts_cool}])
over $\tau_{NTSI}$ (eq.~[\ref{e:ti_ntsi:rho}]), where the latter is used as a proxy
of dynamical instabilities. Since the cooling timescale does not depend on the spatial
scale, dynamical instabilities can become dominant for small scales ($L=0.1\cdots 1.0$~pc) 
in the WNM ($T\approx 10^4$~K). With increasing density, cooling dominates dynamics, but the
TI is still limited by condition~(\ref{e:ti:rho}).

Dynamical instabilities entail the generation of turbulence --
especially at the high Reynolds numbers in the ISM. The turbulent cascade
leads to a spectrum of density perturbations, populating a wide range
of spatial scales. In other words: instead of proceeding at a fixed
scale, the TI will act on all density perturbations within its parameter range
(see Fig.~\ref{f:tidetail}) at scales smaller than the sound crossing scale simultaneously. 

\subsection{An Evolutionary Sequence\label{ss:evol}}

The strength and importance of the TI is based on the one hand on the fact that 
it grows on the smallest scales first \citep{2000ApJ...537..270B} and thus
can feed on ubiquitous small-scale structures generated by turbulence. On the other
hand, the condensation mode of the TI does not proceed at a fixed scale:
a fluid parcel which gets thermally unstable will contract to smaller scales and thus
``delay'' running into the sound crossing scale condition~(\ref{e:ti:rho}).

To see this, consider the thermal equilibrium 
curve (solid black curve in e.g. Figs.~\ref{f:ntplane} and \ref{f:tidetail}). 
Numerical simulations of the fragmentation of shocked flows in a thermally unstable 
medium demonstrate that the gas generally follows this thermal equilibrium curve
(see Fig.~3 of \citealp{2005A&A...433....1A}, and Fig.~14 of \citealp{2006ApJ...648.1052H}),
so that it can be interpreted as an evolutionary sequence from the WNM to the CNM. 
While the equilibrium curve
is well defined at the high-density end due to extremely short cooling time scales 
(see Fig.~\ref{f:tidetail}), the scatter especially in the thermally unstable regime can be 
substantial and depends largely on the ratio between the dynamical and the cooling 
timescales (Figs.~14 and 15 of \citealp{2006ApJ...648.1052H}).

A fluid parcel\footnote{Strictly, the notion of a fluid {\em parcel} is slightly inappropriate.
As the numerical models consistently show, the density structures tend to be filamentary rather than
round. In that sense, the scale refers to the shortest axis of the filament/sheet.} 
in the WNM/WIM starts out at a scale of e.g. $\sim 1$~pc and a density of a little
under $1$~cm$^{-3}$. The parcel may be compressed due to a shock wave, or  
turbulence may lead to density variations. In any case, the density increases, 
propelling the parcel upwards along the isothermal branch of the equilibrium curve, 
reducing the cooling timescales (Fig.~\ref{f:tidetail}) and thus eventually lowering 
the temperature. The fluid parcel will follow (approximately) the equilibrium 
curve along its sharp left-turn into the thermally unstable regime. Looking at the 
1~pc panel, by then, the fluid parcel will have left the region of thermal instability 
(Fig.~\ref{f:ntplane}). However, the compression above not only increased the density,
but also reduced the scale, so that effectively the thermally unstable regime (blue hashed) 
shifts upwards to higher densities, moving us from the $1$~pc panel to the $0.1$~pc panel in 
Figures~\ref{f:ntplane} or \ref{f:tidetail}.

Once the fluid parcel has entered the thermally unstable regime, it evolves (more or less) along lines of
constant pressure towards higher densities and lower temperatures. In the
condensation mode, this is equivalent to a further shrinking of the fluid parcel's volume,
so that the blue ribbon of thermal instability essentially ``moves'' to smaller scales with the fluid parcel 
on its way along the equilibrium curve. 
This ``co-moving'' or Lagrangian behavior is crucial for the strength and the 
importance of the condensation mode of the TI in the ISM. 

Yet the above statement calls for a caveat. A closer inspection of equation~(\ref{e:ti:rho}) reveals the
following. Assuming that the fluid parcel evolves under constant pressure $P_0$ 
-- actually, for dynamical times substantially longer than thermal timescales, the fluid parcel will
follow approximately the thermal equilibrium curve (\citealp{2002ApJ...577..768S}; 
and Fig.~\ref{f:ntplane}), which is sub-isobaric
as indicated by $\gamma_{eff}<0$ in the thermally unstable regime --
and with a constant cooling strength $\Lambda_0$, the sound crossing scale condition~(\ref{e:ti:rho}) translates into an
expression for the outer length scale $L_T$ of thermal instability,
\begin{equation}
  L_T < a_0\,n^{-5/2}\label{e:sc_t},
\end{equation}
equivalent to equation~(\ref{e:cond_soundcross}), with the constant 
\begin{equation}
  a_0\equiv 2\pi\,\left(\frac{\gamma\,k_B}{\mu}\right)^{1/2}\,\frac{P_0^{3/2}}{\Lambda_0}.
\end{equation}
A spherical blob of constant mass $m_0$ will contract as
\begin{equation}
  L_V = b_0\,n^{-1/3},\label{e:sc_v} 
\end{equation}
with the constant
\begin{equation}
  b_0\equiv\left(\frac{6\,m_0}{\pi\,\mu}\right)^{1/3}.
\end{equation}
Thus,
\begin{equation}
  L_T \propto L_V\,n^{-13/6},
\end{equation}
meaning that the sound crossing scale (eq.~[\ref{e:cond_soundcross}]) shrinks more 
rapidly by a factor of $\approx n^2$ than the size of the fluid parcel. Even 
a one-dimensional compression $L_V\propto n^{-1}$ would still result in 
$L_T\propto L_V\,n^{-3/2}$. This would move the fluid parcel out of the thermally 
unstable regime. If the fluid parcel contracts ``faster'' 
than the sound crossing time criterion allows, the condensation 
mode will be suppressed, but the isochoric mode of the TI still will be excited. 
This is specifically the case for the black hashed regions (Fig.~\ref{f:ntplane}) 
where thermal timescales are substantially shorter than the dynamical timescales.
The thermal pressure of the fluid parcel can then drop below that of the surrounding medium,
and the resulting waves will lead to further fragmentation
(smaller scales) and compression, triggering again the condensation mode. In that
sense, while the co-moving picture is a simplification, the more realistic 
evolution of a fluid parcel involving isobaric and isochoric modes will lead
to even more fragmentation. 

Eventually, our fluid parcel will reach the sub-isothermal branch of the thermal 
equilibrium curve below $\approx 80$~K, entering the gravitationally dominated 
regime (as discussed earlier, this regime is characterized also by extremely short 
cooling timescales). In the presence of an ``external'' pressure (due to e.g. 
gravity), the sub-isothermal equation of state can lead to further fragmentation. 

However, gravity will not only play a local role. At large scales, global gravity 
can be dominant (Fig.~\ref{f:ntplane}, e.g. 10~pc). Global gravitational modes 
in combination with a finite extent of the cloud can easily lead to the sweep-up
of material {\em at the edges} of the cloud (\citealp{2004ApJ...616..288B}; see
\citealp{2007ApJ...657..870V} and \citealp{2008ApJ...674..316H} for numerical
evidence). Due to the rapid evolution of the {\em local} (thermal and dynamical) instabilities, 
the resulting filaments will have substructure, leading to further fragmentation into cores.

The thermal instability thus offers a short-cut to reach the gravitationally dominated
regime from the warm, isothermal branch (at $T=10^4$~K). Instead of having to compress
the gas isothermally by a factor of $\approx 10^4$ -- corresponding to a Mach number
of $10^2$ -- only a modest compression is needed in the warm isothermal branch 
to move the fluid parcel up the thermally stable equilibrium line and  ``push it over the ledge''. 
This allows star formation in environments without high pressurization and/or deep gravitational 
potentials. 

To summarize, the generation of turbulence due to dynamical instabilities (red hashed regime)
introduces density variations at smaller scales, triggering the thermal instability, which
in turn amplifies the density perturbations. The growth of the thermal instability is eventually
limited by the mass reservoir available, and thus by 
the external pressure \citep{2002ApJ...569L.127K,2002ApJ...580L..51K}. This could take the form
of global gravity, cloud collisions, expanding shells or galaxy mergers. Gravity acts locally
on the high-density seeds, allowing rapid collapse before edge effects of global gravity 
\citep{2004ApJ...616..288B} overrun the local perturbations. 

\subsection{Limitations\label{ss:limitations}}

Obvious limitations of the study include neglecting the effect of structure in the inflows,
a generic cooling curve, the missing time-dependence and neglecting magnetic fields.
Each of these is discussed in turn below.

\subsubsection{The Structure-less Inflows}
Our analysis approaches the problem of structured cloud formation 
from the angle of a ``worst case'' scenario, using the most
unfavorable initial conditions to generate turbulence and substructure, by considering
(as in our numerical work) the large-scale flows to be uniform. Obviously -- as pointed out earlier
\citep{2006ApJ...648.1052H} -- this assumption neglects the possibility that the
flows themselves already might contain substructure.  This however is not problematic,
as our goal is to demonstrate that substructure can arise from uniform flows with
only small, long-wavelength initial perturbations.  
The introduction of structure in the inflows -- specifically a mixture of WNM/CNM instead 
of just WNM -- would make it even easier to form substructure.  Moreover, introducing
substructure in the inflows raises the question of just what form should be assumed.
The most common initial conditions for numerical simulations of cloud formation
comprise density perturbations (e.g. \citealp{2005AIPC..784..318I}), velocity
``noise'' (e.g. \citealp{2007ApJ...657..870V}), or a turbulent velocity distribution
(e.g. \citealp{2007A&A...465..445H}), but the theoretical justification of these assumptions
is unclear.  Moreover, it is far from obvious that
driving mechanisms such as that of a spherical stellar wind, H II region expansion, or 
supernova bubble expansion are highly turbulent and structured in the absence of
interaction with the surrounding medium.  By ignoring substructure in the inflows we
can consider the most general (and unfavorable) cases for generating turbulent substructure.

\subsubsection{Cooling Curve\label{sss:limitcool}}
Our cooling curve at low temperatures is mainly limited by the fact that it does
not include molecular line cooling or the formation of molecules. Molecular cooling
would lower the cooling timescales further, driving the cooling curve more to
isothermality. The formation of H$_2$ lowers the sound speed by a factor of $\sqrt{2}$ 
and the pressure by $2$, thus introducing an additional pressure loss at high densities.
While the thermal equilibrium curve will describe the evolution of a fluid parcel
approximately correctly from the WNM to the CNM, the details of the evolution at
high densities and low temperature are less realistic.

The heating and cooling functions $\Gamma$ and $\Lambda$ were defined for solar
abundances and a Galactic UV background. Reducing the metallicity by e.g. a 
factor of $10$ to values typical for the SMC reduces the cooling strength 
below $T\approx 10^4$~K by approximately the same factor (see Fig.~\ref{f:coolcurve}).
Thus, the blue ribbon in Figure~\ref{f:ntplane} would move up to higher densities.

\subsubsection{Time dependence}
Obviously, the analysis does not include any time evolution (although
a rough idea of the dynamical evolution can be gleaned from following
the thermal equilibrium curve connecting the WNM and CNM, \S\ref{ss:evol}). 
Dynamical effects (turbulence, shocks) are expected to render thermal effects 
(cooling, TI) even stronger, specifically by providing small-scale seeds, and by
allowing the TI to act on many scales simultaneously. This can clearly be seen in the high-resolution
two-dimensional simulations by \citet{2007A&A...465..445H} and \citet{2007A&A...465..431H}.
For a detailed discussion of (fully 
developed) turbulence in combination with the TI, see e.g. \citet{2003LNP...614..213V}
and \citet{2005A&A...433....1A}. The latter authors point out that increasing the level of
turbulence will lower the CNM fraction, an effect depending on the ratio of dynamical over
thermal timescales (see also \citealp{2006ApJ...648.1052H}).

Certain simplifications had to be made, e.g. the Mach number
for the KHI (eq.~\ref{e:ts_KHI}) was set to $1$.
While this is a good approximation for many regimes in the ISM that exhibit trans-sonic
flows, it will fail in cases of extreme shear flow conditions such as HVCs or 
molecular clouds (however, see Fig.~9 of \citealp{2006ApJ...648.1052H}).

\subsubsection{Magnetic Fields}
Magnetic fields are not included. 
In relevance to the flow-driven cloud formation scenario,
they would introduce a threshold criterion
for the KHI \citep{1961hhs..book.....C} and the NTSI \citep{2007ApJ...665..445H}, 
and they may affect the evolution of the TI \citep{1965ApJ...142..531F}.
Critical field strengths and angles for the suppression of the condensation mode
in one-dimensional converging flows have been identified by \citep{2000A&A...359.1124H}. 
Two-dimensional models of flow-driven cloud formation with the magnetic field
perpendicular to the inflows \citep{2008arXiv0801.0486I} demonstrate that while
the fields can suppress the formation of high-density clouds, thermal fragmentation
still occurs, leading to highly filamentary cold HI clouds. These results lighten
somewhat the restrictive low limits on the perpendicular field strengths derived from
one-dimensional arguments \citep{2004ApJ...612..921B}.
On a larger (galactic spiral arm) scale, numerical models by 
\citet{2003ApJ...599.1157K} and \citet{2006ApJ...646..213K} demonstrate the importance of magnetic
fields for channeling gas streams.


%
%
\section{Summary}\label{s:summary}

To form stars, molecular clouds need to fragment
on scales vastly smaller than their overall dimensions. Self-gravity in combination
with supersonic turbulence allows the rapid fragmentation of Jeans-unstable
regions \citep{1981MNRAS.194..809L,1993ApJ...419L..29E,2004RvMP...76..125M}. 
Yet unless the density perturbations in molecular 
clouds are non-linear {\em before} the onset of global gravitational collapse,
global gravity tends to win and ``overrun'' any local perturbation \citep{2004ApJ...616..288B}. 
Thus, a strong fragmentation mechanism is needed {\em during the formation of the 
cloud} to provide {\em simultaneously} the observed turbulence as well as 
the initial density seeds for the rapid {\em local} gravitational
collapse mandated by the observed rapid onset of star formation 
\citep{2001ApJ...562..852H,2002ApJ...578..914H}. 

The flow-driven formation of molecular clouds in large-scale flows
of atomic hydrogen provides a natural fragmentation mechanism due to
a combination of dynamical and  strong thermal instabilities (see references in 
\S\ref{s:motivation}). As 
demonstrated by numerical models, this fragmentation
mechanism does not need to recur on imposed turbulent velocity fields, but arises
from the physical conditions in the WNM/CNM. The rapid fragmentation of the 
shocked flows will form high-density seeds for local gravity to take over,
while gravitational forces on the cloud scale can sweep up material 
into filaments \citep{2008ApJ...674..316H}. 

The above outline of the connection between cloud formation and the initial
conditions for star formation mainly rests on evidence from numerical simulations. In
the present study we discussed the timescales of the instabilities
dominating the evolution of shocked flows. Guided
by our earlier work, we identify four instabilities, namely two condensation
mechanisms (gravity and thermal instability) and two fluid instabilities
(NTSI and KHI). We determine the parameter regime in density, temperature and
scale in which each of the instabilities is expected to dominate. 
These are our findings and their implications:

\begin{enumerate}
  \item The (dynamical and thermal) instabilities leading to the rapid fragmentation
        of shocked flows dominate on small scales for reasonable parameters of the
        WNM and CNM (Fig.~\ref{f:ntplane}).  
  \item Cooling timescales (of $\approx 1$~Myr
        in the thermally unstable regime) are generally substantially shorter than
        gravitational timescales for that same density and temperature regime
        (Fig.~\ref{f:tipair}). Thus, thermal fragmentation dominates gravitational
        fragmentation during cloud formation.
  \item The strength of the condensation mode of the TI derives from two effects,
        namely that it grows on the smallest scales 
        first, and that it is essentially a co-moving
        (or Lagrangian) mode, i.e. the unstable region shrinks, thus keeping
        ``longer'' below the sound crossing scale $\lambda_c$ (eq.~[\ref{e:ti:rho}]). Feeding on turbulent density
        perturbations below $\lambda_c$ will allow the TI to grow on many scales simultaneously.
  \item The preference of small scales by the TI entails the preference to form
        sheets and filaments: the shorter axes are more susceptible to become unstable.
        This is an obvious parallel to the GI, with the -- equally obvious but crucial -- difference
        that the GI will act globally {\em and} locally.
  \item The fragmentation due to thermal effects is crucial for the formation of
        small-scale, high-density perturbations {\em during the formation of the 
        molecular cloud}, to provide the seeds for rapid {\em local} gravitational
        collapse before global gravity dominates the dynamics completely. 
  \item The all-important role of the TI due to the intermittent nature of the turbulence
        driven by dynamical instabilities allows an early fragmentation of the 
        converging flows,
        before (local) gravity can take over. This supports numerical findings that 
        the (molecular) core mass spectrum might be set to some extent
        \citep{2008arXiv0801.2257D} early on during cloud formation
        by thermal fragmentation \citep{2007A&A...465..445H,2008ApJ...674..316H}.
\end{enumerate}

\acknowledgements 
We thank the referee for a critical and timely report. 
The arguments presented lean heavily on
the evidence from numerical simulations performed at the National Center for
Supercomputing Applications (AST 060031). FH is supported by the University
of Michigan and NASA grant NNG06GJ32G.
This work has made use of NASA's Astrophysics Data System.

%
%


\end{document}